# Excitation Function of $^{12}$C(p,p$_o$) for E$_p$=16 ÷ 19.5 MeV measured by MSS with resolution ~10 keV.

A. Gafarov[1*]

[1]Nuclear Reactions Laboratory, INP of Science Academy of Republic Uzbekistan

[*]email: anatxagor@gmail.com

The elastic scattering $^{12}$C(p,p$_o$) Excitation Function (EF) was measured in the energy range $E_p$ =16 ÷ 19.5 MeV with resolution ~10 keV by means of innovative approach - **MSS** - the *Method of Spectra Superposition* at the 14-angle Magnetic Spectrograph (Apelsin) at beam with an energy spread of ~200 keV from U-150 cyclotron of INP Ulugbek (Tashkent) of Science Academy of Uzbekistan [1-8]. The obtained EF has a reach structure of anomalies in precise agreement with thresholds & levels data [2,9-12]. Measurements were done on the $^{12}$C self-supporting target with thickness 13 mg/cm$^2$ and an area of 1 cm$^2$ in the center of Apelsin. The 20-step Energy Moderator controlled the proton energy, providing a 3.5 MeV wide $E_p$ interval with no readjustment of cyclotron and all the ionic-optics on the 41-meter-long beam-pipe. Energy resolution of Apelsin for protons is better than 5 keV along all focal plane, where particle-product were detected by specially designed coordinate-sensitive gas mixture MWPC with two particle counters in coincidence. The acquisition system was based on IBM PC online with a fast CAMAC branch (custom designed at St-Petersburg INP) with fast TDC modules and other fast electronics providing immediate selection & accumulations of events from MWPCs. The NMR-monitor system stabilized spectrographs magnetic field to 3ppm [1,9]. The EF of $^{12}$C (p,p$_o$) scattering has many resonances that precisely correspond to g.s. and levels of well-known product-nuclei. But 80% of anomalies didn't have any explanation. A new concept is proposed to explain all unrecognized anomalies. Two expressions *Writing I* and *Writing II* proposed to explain the unknown peaks which corresponds to population of A13 phase volume states. The results of the proposed explanation look very promising (Fig. 13). Concept of the Combinative Isobaric Resonances (CIRs) proposed.

## 1. Introduction

The idea to consider each beam-particle as the **Dirac's δ-function** is not just promising, but a highly productive one. It makes fully real the high-energy resolution studies regardless of the wide beam energy-spread.

This idea was the basic for the ***Method of Spectra Superposition – MSS***, that allowed to measure EF with a resolution of ~10 keV at a beam with a ~200 keV energy-spread [1-9]. The **MSS** made possible to combine high energy-resolution detection with an ordinary accelerator like cyclotron.

Several key things provided success to **MSS** and each of them was the essential component.

1) The 14-angle magnetic spectrograph **Apelsin** with high energy-resolution for charged particles, less than 5 keV for 20 MeV protons.
2) The electronic fast coordinate-sensitive MWPC-detection-system (with resolution less than 0.2 mm) embedded into the focal planes of **Apelsin**. (Before detection was based on nuclear photo-emulsions).
3) Acquisition system based on IBM PC online controlling a custom made fast CAMAC-bus designed at St-Petersburg (Leningrad) PNPI.
4) The *Al*-foil beam-energy moderator for rapid energy control with no readjustment of cyclotron and all ionic-optics on the 41-meter long beam pipe (10 lenses and magnets). Just imagine how much beam time it saved!
5) The NMR-monitoring of magnetic field at gaps of **thoroidal magnet of Apelsin**.
6) The final thing was to move the main Acquisition system CAMAC-crate to **Apelsin's** detection sector, what made the TDC input cables as short as possible (before they were about 200m long each, slowing down the front edges of fast-timing pulses).

It immediately resulted in a display of sharp structures on the elastic-process peaks from (**p,$^{12}$C**)-scattering.

## 2. Goals of the measurement

Major goal of this experiment was to prove the **MSS** idea at high energy-resolution magnetic spectrograph in combination with cyclotron and also get the Excitation Function precise data that don't exist yet.

The (**p**,$^{12}$**C**)-process was chosen as the one of the most astrophysical interest. Also, precise data were absent for proton energy-range $E_p$ =16 ÷ 19.5 MeV. A self-supporting thin $^{12}$C target of 13 mg/cm² with an aperture of 1 cm² was used in the study.

## 3. Experimental setup

**3.1.** Detection system

All things were circling around the 14-sector magnetic spectrograph **Apelsin** as the most advanced detecting system for protons energy up to 28 MeV and alpha-particles with energy up to 40 MeV (Fig. 1).

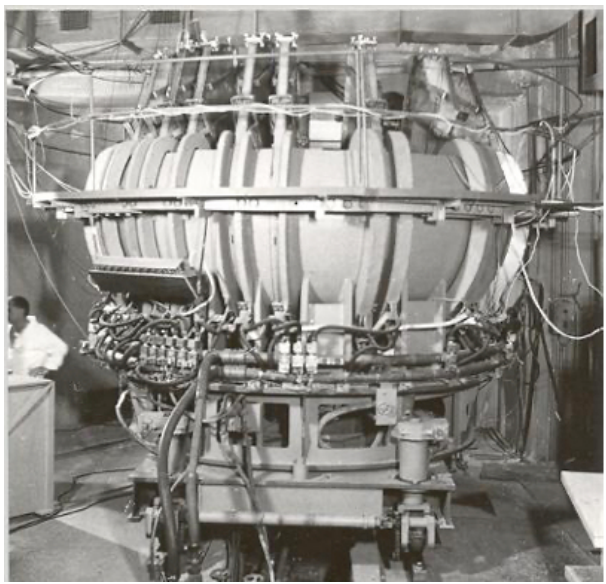

Fig. 1. The high energy-resolution 14-angle magnetic spectrograph **Apelsin (Orange)**.

Being designed for nuclear-emulsion plates, **Apelsin** required a deep redesign of the registration system to provide real-time control of the things detected. Otherwise several days could be spent for emulsion irradiation and then one discovers something was wrong, beam-time lost, and all data compromised.

That's why for the focal planes a special particle detection system, based on coordinate gas proportional chambers –MWPC, was custom designed together with scientists from PNPI (former Leningrad INP, Gatchina, USSR), Fig. 2 – one of two counters against each other. The one-coordinate MWPC has in common gas box two counters in coincidence, where the upper anode (red), 20 $\mu м$ **Au**-plated **W**-micro-wire, was positioned into the focal axis of the magnetic spectrograph analyzing segment. Incident particles, passing through the anode counting area, produce the Coulomb micro-burst **(+)** of positive gas-ions. Each Coulomb-burst is a source of the two soliton waves in the near located flat coordinate electromagnetic delay line (spiral EDL, 50 Ω, 580 ns) along the entire length of anode wire – 300 mm. The screened EDL (blue) has a signal-window (green).

The time difference, in the signals 1 and 3 from opposite ends of EDL, says what is coordinate of the Coulomb-burst. Each EDL on the ends has FET-based preamplifiers. Then timing signals go outside the MWPC into the main preamplifiers combined with the CF-discriminators. The resulting time jitter was less than 100 ps what corresponds to a coordinate resolution better than 0.1 mm along the focal plane axis.

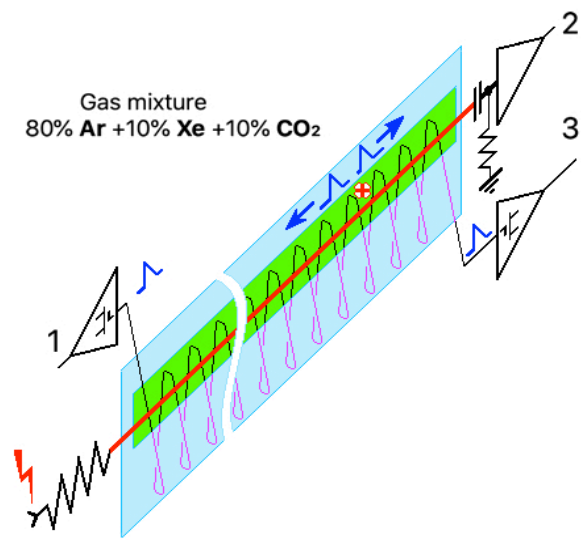

Fig. 2. Sketch of the **upper** of two counters in the MWPC. Next counter looks like mirror-reflected, but its anode and signal-window (green) are near the bottom.

The MWPC tests and calibrations, under **X**-rays of sharp-collimated $^{55}$**Fe** (5.9 keV) and β from $^{106}$**Ru**-sources, resulted in coordinate-resolution ∼ 0.1mm. Also, all tests and calibrations were done under α-particles from the sharp-collimated $^{238}$**Pu** what gave resolution ∼ 0.2mm. Gas mixture in parts 8Ar+Xe+CO₂ was flowing under atmospheric pressure continuously.

The MWPC-chamber was separated from the vacuum space with 7$\mu м$ aluminized mylar film. Under the atmosphere pressure, the gas mixture was slowly non-stop flowing, from the high-pressure gasholder through the MWPC volume then back into atmosphere via oil nipple during all beam time [1,3,10-13].

### 3.2. Experimental area

The setup on Fig. 3, included a cyclotron U-150 with $E_p \leq 22$ MeV, the 41-meter long beam pipe with 3 magnetic prisms, 10 quadrupole lenses, and also the 14-angular Magnetic Spectrograph "Apelsin" as product-particle detector. The FWHM of "Apelsin" is better than 5 keV for protons (15÷20 MeV) [1,3]. The NMR-monitoring system stabilized the "Apelsin" magnetic field to 3ppm.

### 3.3. Acquisition system

The MWPC both counters work in coincidence (fast signals from both anodes and both cathode delay lines). The coincidences "select" the gliding entrée angle 36° of the incoming detected product-particles and insured really deep suppression of the background radiation (Fig.4).
The anode wire of the leading "upper" counter was positioned onto the focal plane of section #7 at the gap of the toroidal yoke of "Apelsin"  The anodes time-jitter was ≤ 10 ps.

To provide even deeper background radiation suppression, a module [FD]+PMT-based fast-plastic scintillator counter was introduced as monitor of the beam real-time micro-bunches, via the $\gamma$-bursts from the luminophore-covered gate-slit at the reaction-chamber entrance of "Apelsin"  (the dark blue filled area - vacuum). For whole acquisition system, the U-150 **HF**-generator's signal via [BS] and [FD]+PMT chain were the basic in generation of main trigger.

As shown on Fig. 4, product-particles, from the target at the reaction chamber center, fly away through entrance slit-system into analyzing sectors (under angle of 16.3° to the beam direction). Magnetic field in gaps of each two sections of the toroidal core (yoke gaps) transforms the product-beam into the angle-dispersed spectrum, focused on the anode wire of upper (leading) coordinate counter of MWPC.

Delay time difference in signals from both ends of MWPC-cathodes twice marks trajectory-coordinates of each registered particle with a jitter of ≤ 0.2 mm on the focus plane and behind it to select right entrée angles.
A specially designed ring-moderator (with 20 windows) rapidly changed $E_p$ in 20 jumps from 19.5 MeV to 16 MeV with no readjustment of cyclotron and 14 elements of the ion optics on the beam pipe.

The acquisition system was based on IBM PC that controlled the moderator and a CAMAC-branch with a special set of the nanosecond-fast modules (designed at St-Petersburg PNPI) including fast TDC and other fast electronics.
Such an architecture provided immediate selection & collection of events from MWPCs and Faraday cup with the [I/F]-convertor of the beam-current into frequency.

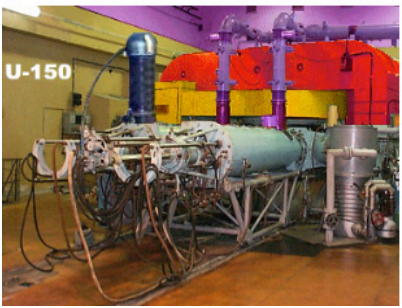
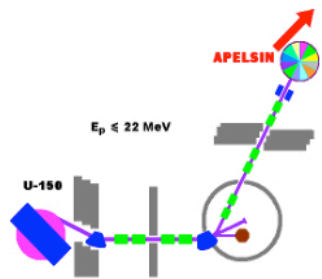
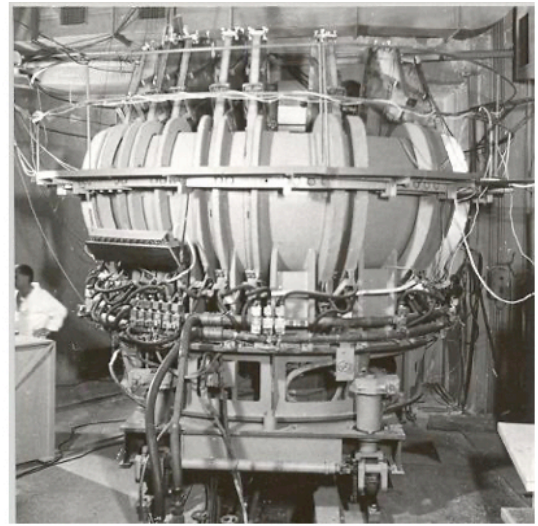

Fig 3.      Cyclotron U-150.         Sketch of experimental area.     14-angular Magnetic Spectrograph "Apelsin"

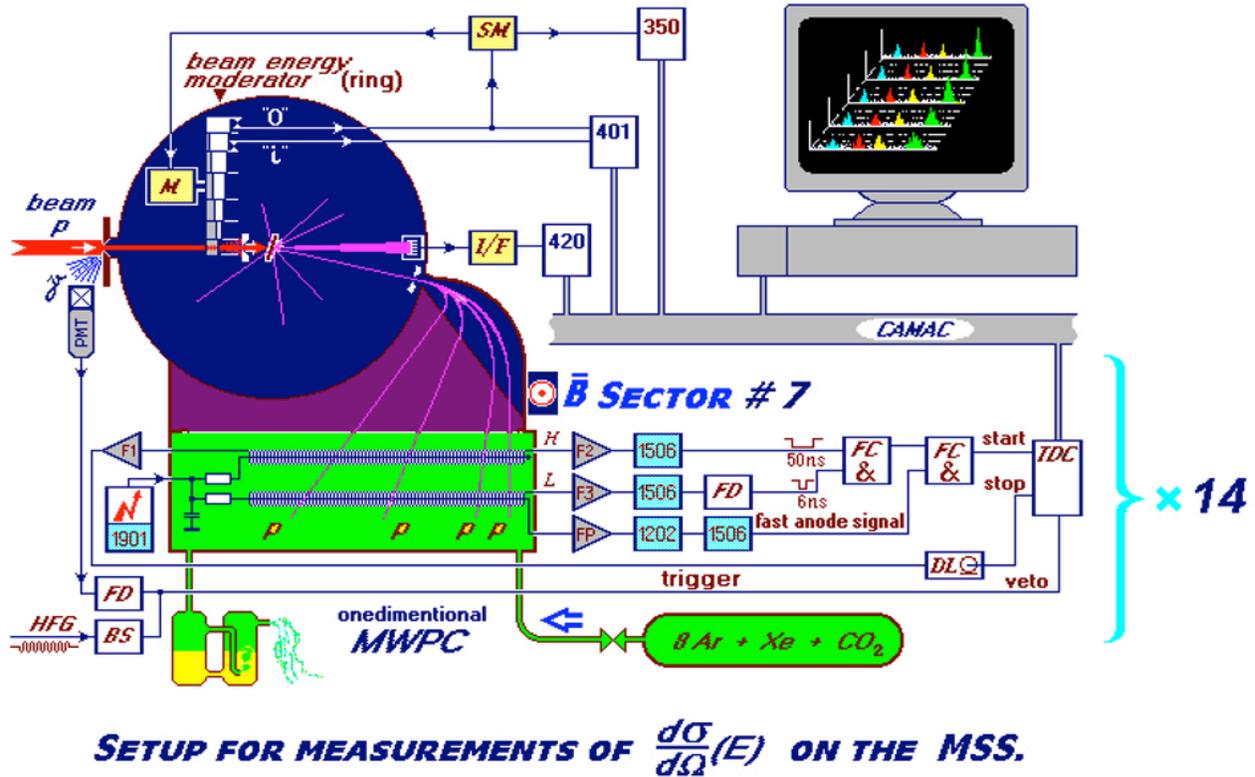

Fig. 4

## 4. The MSS-measurement procedure

The **MSS**-setup works in cycles of 20 expositions of target under 20 different beam-energies.

First of all, one chose the initial beam-energy $E_o$ what means one will have an energy range of ~ 3.5 MeV wide $\{E_{20} \div E_o\}$, where $E_{20}= E_o – 3.5$ MeV.

Then one chose under what angle $\Theta_{Lab}$ to make measurements.

**1)** The exposition "0" starts from $E_o$ – the moderator in position "0" – open window.

Acquisition system starts measurement of first typical spectrum of particle-products with elastic and inelastic (if any) processes (Fig. 5 gray filled, or Fig. 6, group (R) spectrum with $A_o$, $B_o$, $C_o$, $H_o$). Measurement is over when the needed statistics under the needed peak of process of interest ($A_o$, $B_o$, $C_o$, $H_o$) is accumulated.

IBM PC memory stores the measured spectrum with ($A_o$, $B_o$, $C_o$, $H_o$) and $Q_o$ total integrated beam charge.

No stop of cyclotron!

**2)** Exposition "1" starts when moderator installs into the beam thickness #1 plate. One will have beam-energy $E_1$ in just a second without touching whole beam-pipe optics or cyclotron.

Acquisition system starts measurement of next typical spectrum of particle-products with elastic and inelastic (if any) processes (Fig. 6, group (R) spectra with $A_1$, $B_1$, $C_1$, $H_1$. Measurement is over when the needed statistics under needed peak of process (of interest $A_1$, $B_1$, $C_1$, $H_1$) is accumulated in terms of $Q_1 = K_1 \times Q_o$.

Computer's memory stores the measured spectrum with ($A_1$, $B_1$, $C_1$, $H_1$) and $Q_1$ total integrated beam charge…

**20)** Last exposition "19" results in products-spectrum with peaks $A_{19}$, $B_{19}$, $C_{19}$, $H_{19}$ and computers memory stores it together with $Q_{19} = K_{19} \times Q_o$.

First cycle done and one can repeat it if statistics or any other cause requires (something wrong was found…).

The fast TDC has time-resolution ~ 10 picoseconds, what allowed to use 1024 channels of coordinate scale with final resolution ≤0.2 mm along 300 mm of the "Apelsin" focal-plane (MWPC upper counter anode wire).

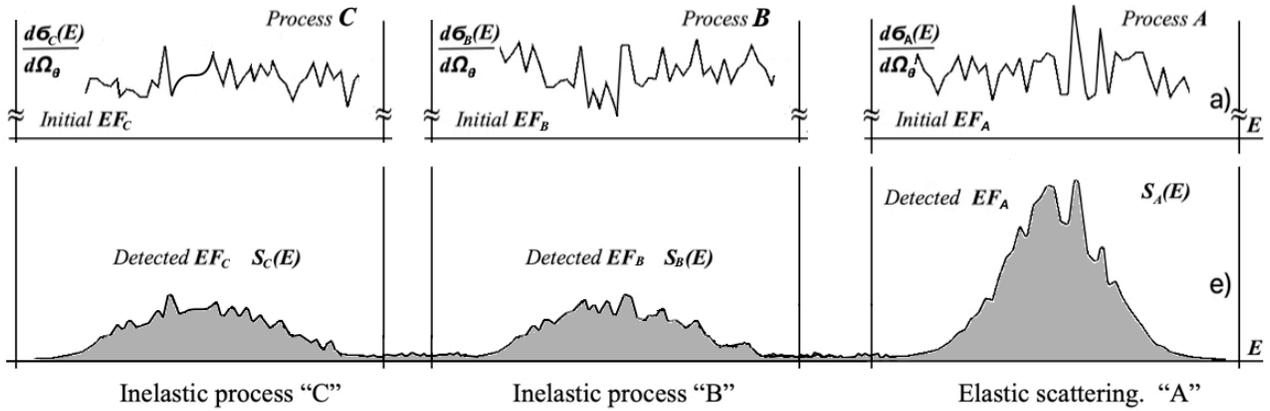

Fig. 5. Excitation functions of interactions **A, B,** and **C** with typical HR-detected spectrum of product-particles.

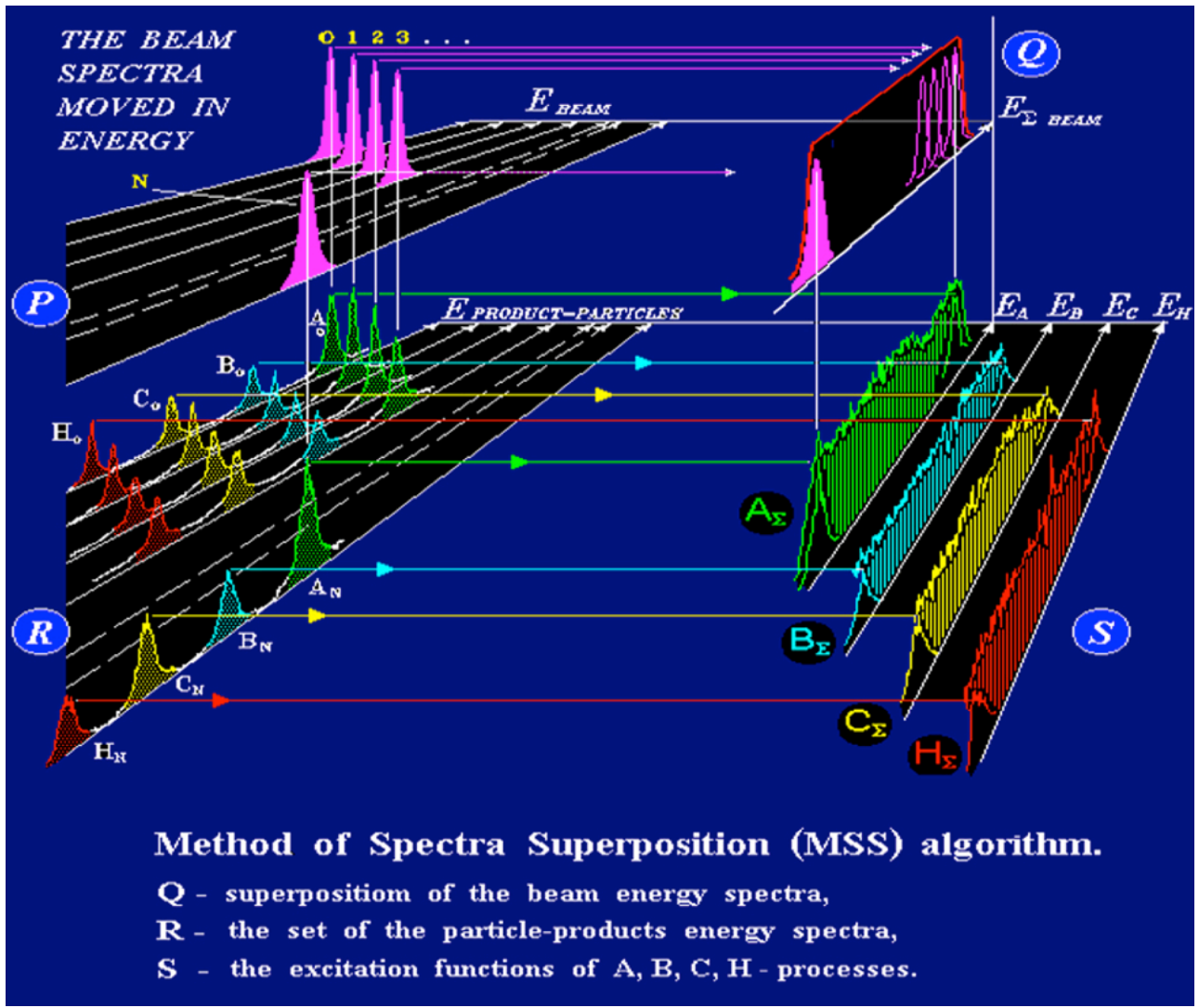

Fig. 6. The **MSS**-algorithm: beam-particle peaks (**P**) and beam superposition spectra (**Q**), interaction peaks (**R**) ($A_i$, $B_i$, $C_i$, $H_i$) with superposition set (**S**) of the elastic scattering **A** and inelastic reactions **B**, **C**, and **H**.

In the real experiment, the 512-channel scale was used. Integral and differential nonlinearity of the coordinate scale finely were 0.4% and 1.6% correspondingly.

Tilt of the product-particle trajectory to the focal plane of "Apelsin" was ~ 36° (for the MWPC upper counter anode axis).

## 5. Data processing

Real beam-particle superposition spectrum (Q on Fig.4) is not flat and, it requires correction. That's why every peak $A_i$ from the $A_\Sigma$ was processed first by moving-average smoothing 11 times among 5 bins, then each resulted peak was fitted to a set of main $MG_i$ (central) + two (left and right side) background Gaussians (*MINUIT*).

The MWPC also has its own "non flat" registration effectivity – $CE_i$ along the anode wire axis. It was studied for protons, α, β, and X-rays. The normalized $CE_i$ curve is - $NCE_i$ -it must be taken into account.

The result – the **normalized to 1** sum of all 20 smoothed main peaks, treated by the $NCE_i$ curve, gives the final correction curve "UMDCF".

Almost the same result is reachable if one applies the *Au* gold leaf target.

Such a special processing by 10-parameter *MINUIT* is applied because of the delta-electrons influence especially around spots of high-intensity (peaks on the focal plane – main anode of MWPC).

**The elastic EF extraction** was done on the first step by subtracting $MG_i$ from each $A_i$ peak (Fig.6, (S)-group):

$$EF_i = A_i - MG_i \qquad , (8)$$

Then the correlation controlled oversewing (sum) of all 20 $EF_i$-peaks on one 512-bin scale gives the raff elastic **EF** which, after correction by UMDCF, becomes the final **elastic $EF_i$** statistics.

The last one was transposed to the relativistic momentum scale by (9) with some of "Apelsin" and MWPC constants:

$$S[pc_i] = EF_i \frac{[80+1.25(X_i + C_{LCN} + CL_{CHI} + L_{VB} + Y/tg(36°))]}{A_{CMB} \, W_{oCN}} , (9)$$

Where $X_i$ -is transposition of TDC-scale onto real focal plane, $A_{CMB}$ -is fraction of spectral line accepted by anode counter in the section (#7) of spectrograph and $W_{oCN}$ – solid acceptance angle of this spectrograph analyzing sector (#7) in steradians.

Then $S(pc_i)$ was transposed to the projectile energy-scale, where the binning was calculated in decimal metric scale, what resulted in $S(E_j)$.

Real cross-section, as a function of projectile-energy $E_j$, was calculated in [mb/sr] via (10):

$$\frac{d\sigma(E_j)}{d\Omega_{LAB}} = \frac{S(E_j) \, A}{h \, S_T \, N_A \left[ \frac{U_{sp}}{\Delta J} \frac{Q_{int}}{1.602 \cdot 10^{-19} Q} \frac{10^{-6} Q}{991} \right]}, (10)$$

with $S(E_j)$ -the product-particle statistics on energy-scale, $A$- atomic weight of the target element, $h$ -target thickness [mg/cm²], $S_T$ – area of the beam-spot on target [cm²], $N_A$ - Avogadro's number, $U_{sp}$ -coefficient of the used fraction of total spectrum of event statistics, $\Delta J = J_{max} - J_{min}$ - number of the used channels on spectrum, $Q_{int}$ -total number of beam-integrator pulses.

This processing finally results in several files:
- 1 -of reaction statistics on *pc*-scale of rel.momentum;
- 2 -of reaction statistics on scale of $E_{beamLAB}$ ;
- 3 file of $d\sigma(E_{beamLAB})/d\Omega_\theta$ for $\{E_{20} \div E_o\}$ energy range of 3.5 MeV.

## 6. Computer simulation

To ensure correctness of data processing, a computer simulation was performed in which the artificial EF was first generated by URAND software. Then it was processed by convolutions to get 20 separate peaks with piloting displayed EF fractions on them.
After that whole MSS-algorithm was applied. The simulation showed stability of the results – the artificial EF was restored with errors less than 15% when the centroids of "beam" peaks where totally unknown or got shifted (magnetic field or accelerator's energy temporary change…).
In case of the known medians of "beam"-peaks, the residual errors were much smaller than 5%.

## 7. First excitation function by MSS for the elastic scattering $^{12}C(p,p_o)$

The **EF** of ($p$, $^{12}C$), measured in the energy range of (16 MeV÷19.5 MeV) by the MSS-approach (Fig.7), showed a resonance reach structure newer observed before in this energy region [1-3, 19-22]. Comparison with [19-22] easily displays it (set of Δ on the Fig.7).

Some of peaks are well known and precisely fit the energy thresholds & levels from large amount of data [19-21]. It's good time to check again all the excited levels data related to the target-nucleus and also all that concerns the products, even no one of the inelastic processes measured in this experiment.

## Reading 0.

$$E_{p\,RF} = E^*_{LNP} + E^*_{RC} + \Delta M_{OS} - \Delta M_{PS}$$

proton rest frame energy / level energy of nucleus-product / mass excess of output system / mass excess of primary system

$\Delta M_{OS} = \Delta M_{NP} + \Delta m_{RC}$       $\Delta M_{PS} = \Delta M_{TN} + \Delta m_{BP}$

$\Delta M_{NP}$ – mass excess of nucleus-product
$\Delta m_{RC}$ – mass excess of residual product-cluster(s) or/and particle(s)
$\Delta M_{TN}$ – mass excess of target-nucleus
$\Delta m_{BP}$ – mass excess of beam-particle

**Writings I and II** for all known product-nuclei in the **IS** lead to the **Reading 0**

| | | |
|---|---|---|
| ← 7.2748 | 3 $^4He^*$ | (+ p) |
| ← 7.3665 | $^8Be^*$ | (+ $^4He$ + p) |
| ← 7.5516 | $^9B^*$ | (+ $^4He$) |
| ← 9.2388 | $^5Li^*$ | (+ 2 $^4He$) |
| ← 9.3306 | $^8Be^*$ | (+ $^5Li$) |
| ← 12.3706 | $^5Li^*$ | (+ $^8Be^{1*}$) |

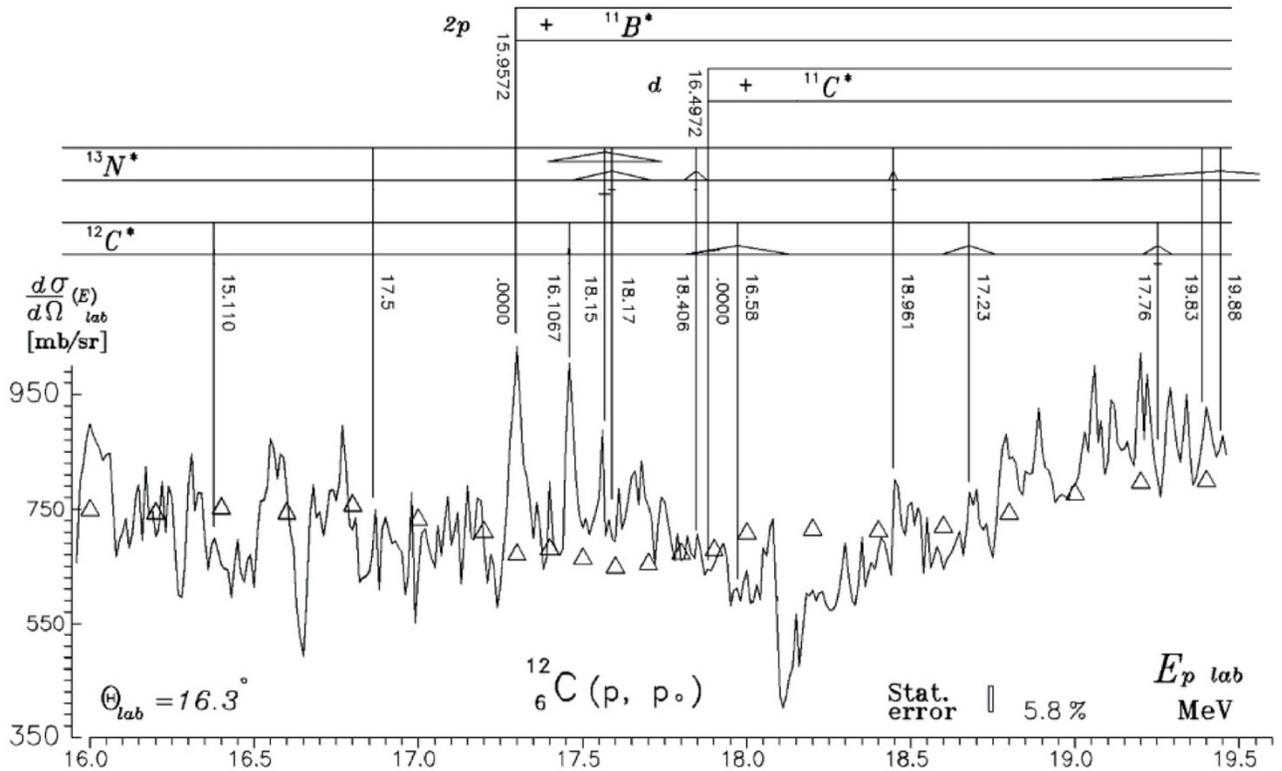

Fig. 7. Excitation function of $^{12}C(p, p_o)$ measured in the range 16 ÷ 19.5 MeV by MSS-approach at magnetic spectrograph "Apelsin" and U-150 cyclotron beam with $E_p$ =19.5 MeV with energy-spread ~200 keV. Δ - from W. W. Daehnick and R. Sherr. Phys. Rev. 133, B934

Well-known expression allows to calculate in the CM-frame the minimal energy that brings different product-nuclei in excited or g.s. (according to the charge conservation law) so, there is the "READING 0":

$$E_{Pcm} = E^*_{LNP} + E^*_{RC} + \Delta M_{OS} - \Delta M_{PS}, \quad (11)$$

$$\Delta M_{OS} = \Delta M_{NP} + \Delta m_{RC},$$

$$\Delta M_{PS} = \Delta M_{TN} + \Delta m_{BP},$$

where $E_{Pcm}$ -is the projectile's energy in CM-frame, $E^*_{LNP}$ -is the excitation energy of the product-nucleus (level energy), $E^*_{RC}$ -is the excitation energy of the residual cluster, $\Delta M_{OS}$-is the mass-excess of the output system, $\Delta M_{PN}$ -is the mass-excess of the product-nucleus, $\Delta m_{RC}$ -is the mass-excess of the residual cluster, $\Delta M_{PS}$ -is the mass-excess of primary system, $\Delta M_{TN}$ -is the mass-excess of the target-nucleus, $\Delta m_{BP}$ -is the mass-excess of the beam-particle.

All anomalies in the EF are proportional to $\sigma$ (rates) of population of all energy-accessible "cells" in the phase volume of $p + {}^{12}C$ interaction, even no other than elastic product-$p$ was detected.
It works on the schematic:

Primary system → Populated system → decay into primary system
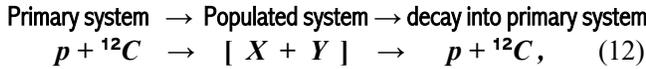
$$p + {}^{12}C \rightarrow [X + Y] \rightarrow p + {}^{12}C, \quad (12)$$

First candidates, to appear as resonances in the EF, are the levels of ${}^{12}C^*$ that protons within 16 MeV – 19.5 MeV can populate:
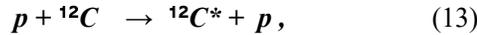
$$p + {}^{12}C \rightarrow {}^{12}C^* + p, \quad (13)$$

The ${}^{12}C^*$ diagram of energy levels is right over the EF on Fig. 7, and the corresponding peaks are 15.110 MeV, dazzling 16.1067 MeV, 16.58 MeV, 17.23 MeV, and 17.76 MeV. The triangle bars represent each level known width and error bars are right below, they cross verticals. The brightest peak in the EF middle corresponds to:

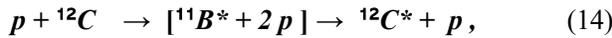
$$p + {}^{12}C \rightarrow [{}^{11}B^* + 2p] \rightarrow {}^{12}C^* + p, \quad (14)$$

It precisely seats under the vertical ${}^{11}B$ g.s. line, which is 15.9572 MeV in the rest frame.

Population of ${}^{13}N^*$ brings some bright resonances:

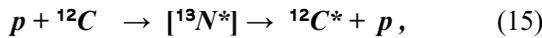
$$p + {}^{12}C \rightarrow [{}^{13}N^*] \rightarrow {}^{12}C^* + p, \quad (15)$$

Here they are: 17.5 MeV, 18.15 MeV, also 18.17 MeV, 18.406 MeV, 18.961 MeV, 19.83 MeV and 19.88 MeV.

Also, ${}^{11}C$ g.s. ( + $d$ ) is right in the middle of. Fig.7.

Some other products do not have any level or g.s. in this proton energy range of 16 MeV– 19.5 MeV. Here below they are:

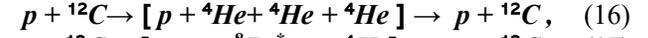
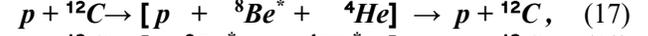
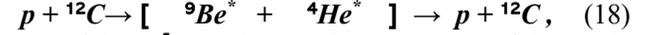
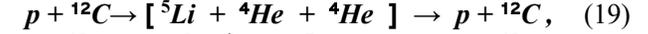
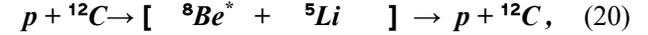

$$p + {}^{12}C \rightarrow [p + {}^{4}He + {}^{4}He + {}^{4}He] \rightarrow p + {}^{12}C, \quad (16)$$
$$p + {}^{12}C \rightarrow [p + {}^{8}Be^* + {}^{4}He] \rightarrow p + {}^{12}C, \quad (17)$$
$$p + {}^{12}C \rightarrow [{}^{9}Be^* + {}^{4}He^*] \rightarrow p + {}^{12}C, \quad (18)$$
$$p + {}^{12}C \rightarrow [{}^{5}Li + {}^{4}He + {}^{4}He] \rightarrow p + {}^{12}C, \quad (19)$$
$$p + {}^{12}C \rightarrow [{}^{8}Be^* + {}^{5}Li] \rightarrow p + {}^{12}C, \quad (20)$$

**General view on Fig. 7 looks strange. Almost all other unrecognized peaks (80%) are bigger than 3σ, but what they are?** The statistical error is less than 6%. They must be something that matters!

## 8. The source of additional resonances

**Let's, just for fun, check all energy accessible combinations of nuclei + residuals which give A=13, temporarily ignoring the charge conservation…**

One can use the same expression (11) in *Writing I* form to calculate positions of levels and ground states.
*Writing I*:

$$E_{Pcm} = E^*_{LIN} + E^*_{RC} + \Delta M_{IS} - \Delta M_{PS}, \quad (21)$$

$$\Delta M_{IS} = \Delta M_{IN} + \Delta m_{RC},$$

$$\Delta M_{PS} = \Delta M_{TN} + \Delta m_{BP},$$

where $E_{Pcm}$ -is the projectile's energy in CM-frame, $E^*_{LIN}$ -is the excitation energy of the **initiated** nucleus (level energy), $E^*_{RC}$ -is the excitation energy of the residual cluster, $\Delta M_{IS}$-is the **initiated** system mass-excess, $\Delta M_{IN}$ -is the mass-excess of the **initiated** nucleus, $\Delta m_{RC}$ -is the mass-excess of the residual cluster, $\Delta M_{PS}$ -is the mass-excess of primary system, $\Delta M_{TN}$ -is the mass-excess of the target-nucleus, $\Delta m_{BP}$ -is the mass-excess of the beam-particle.

**Just for fun, (13) is going to split into isobaric combinations below:**
primary system, initiated nucleus & cluster

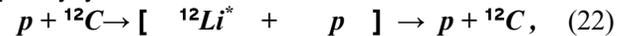
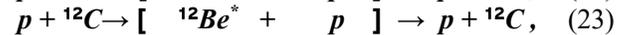
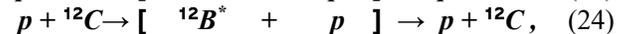
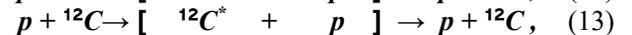
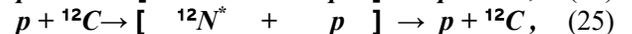
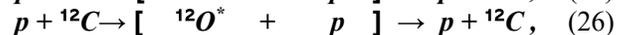
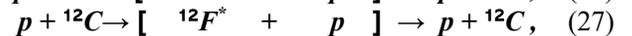
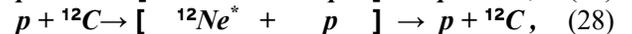

$$p + {}^{12}C \rightarrow [{}^{12}Li^* + p] \rightarrow p + {}^{12}C, \quad (22)$$
$$p + {}^{12}C \rightarrow [{}^{12}Be^* + p] \rightarrow p + {}^{12}C, \quad (23)$$
$$p + {}^{12}C \rightarrow [{}^{12}B^* + p] \rightarrow p + {}^{12}C, \quad (24)$$
$$p + {}^{12}C \rightarrow [{}^{12}C^* + p] \rightarrow p + {}^{12}C, \quad (13)$$
$$p + {}^{12}C \rightarrow [{}^{12}N^* + p] \rightarrow p + {}^{12}C, \quad (25)$$
$$p + {}^{12}C \rightarrow [{}^{12}O^* + p] \rightarrow p + {}^{12}C, \quad (26)$$
$$p + {}^{12}C \rightarrow [{}^{12}F^* + p] \rightarrow p + {}^{12}C, \quad (27)$$
$$p + {}^{12}C \rightarrow [{}^{12}Ne^* + p] \rightarrow p + {}^{12}C, \quad (28)$$

Set (14) is going to split into isobaric combinations:

$$p + {}^{12}C \to [{}^{11}Li^* + 2p] \to {}^{12}C^* + p, \quad (29)$$
$$p + {}^{12}C \to [{}^{11}Be^* + 2p] \to {}^{12}C^* + p, \quad (30)$$
$$p + {}^{12}C \to [{}^{11}B^* + 2p] \to {}^{12}C^* + p, \quad (14)$$
$$p + {}^{12}C \to [{}^{11}C^* + 2p] \to {}^{12}C^* + p, \quad (31)$$
$$p + {}^{12}C \to [{}^{11}N^* + 2p] \to {}^{12}C^* + p, \quad (32)$$
$$p + {}^{12}C \to [{}^{11}O^* + 2p] \to {}^{12}C^* + p, \quad (33)$$
$$p + {}^{12}C \to [{}^{11}F^* + 2p] \to {}^{12}C^* + p, \quad (34)$$
$$p + {}^{12}C \to [{}^{11}Ne^* + 2p] \to {}^{12}C^* + p, \quad (35)$$

Set (15) is going to split into isobaric combinations:

$$p + {}^{12}C \to [{}^{13}Li^*] \to {}^{12}C^* + p, \quad (36)$$
$$p + {}^{12}C \to [{}^{13}Be^*] \to {}^{12}C^* + p, \quad (37)$$
$$p + {}^{12}C \to [{}^{13}B^*] \to {}^{12}C^* + p, \quad (38)$$
$$p + {}^{12}C \to [{}^{13}C^*] \to {}^{12}C^* + p, \quad (15)$$
$$p + {}^{12}C \to [{}^{13}N^*] \to {}^{12}C^* + p, \quad (39)$$
$$p + {}^{12}C \to [{}^{13}O^*] \to {}^{12}C^* + p, \quad (40)$$
$$p + {}^{12}C \to [{}^{13}F^*] \to {}^{12}C^* + p, \quad (41)$$
$$p + {}^{12}C \to [{}^{13}Ne^*] \to {}^{12}C^* + p, \quad (42)$$

Set (16) is going to split into isobaric combinations:

$$p + {}^{12}C \to [p + {}^{4}He + {}^{4}He + {}^{4}He] \to p + {}^{12}C, \quad (16)$$
$$p + {}^{12}C \to [p + {}^{4}Li + {}^{4}He + {}^{4}He] \to p + {}^{12}C, \quad (43)$$
$$p + {}^{12}C \to [p + {}^{4}Li + {}^{4}Li + {}^{4}He] \to p + {}^{12}C, \quad (44)$$
$$p + {}^{12}C \to [p + {}^{4}Li + {}^{4}Li + {}^{4}Li] \to p + {}^{12}C, \quad (45)$$

Set (17) is going to split into combinations:

$$p + {}^{12}C \to [p + {}^{8}Be^* + {}^{4}He] \to p + {}^{12}C, \quad (17)$$
$$p + {}^{12}C \to [p + {}^{8}Be^* + {}^{4}Li] \to p + {}^{12}C, \quad (46)$$
$$p + {}^{12}C \to [p + {}^{8}Li^* + {}^{4}He] \to p + {}^{12}C, \quad (47)$$
$$p + {}^{12}C \to [p + {}^{8}Li^* + {}^{4}Li] \to p + {}^{12}C, \quad (48)$$
$$p + {}^{12}C \to [p + {}^{8}B^* + {}^{4}He] \to p + {}^{12}C, \quad (49)$$
$$p + {}^{12}C \to [p + {}^{8}B^* + {}^{4}Li] \to p + {}^{12}C, \quad (50)$$
$$p + {}^{12}C \to [p + {}^{8}C^* + {}^{4}He] \to p + {}^{12}C, \quad (51)$$
$$p + {}^{12}C \to [p + {}^{8}C^* + {}^{4}Li] \to p + {}^{12}C, \quad (52)$$

Set (18) is going to split into isobaric combinations:

$$p + {}^{12}C \to [{}^{9}Be^* + {}^{4}He^*] \to p + {}^{12}C, \quad (18)$$
$$p + {}^{12}C \to [{}^{9}Be^* + {}^{4}Li^*] \to p + {}^{12}C, \quad (53)$$
$$p + {}^{12}C \to [{}^{9}Li^* + {}^{4}He^*] \to p + {}^{12}C, \quad (54)$$
$$p + {}^{12}C \to [{}^{9}Li^* + {}^{4}Li^*] \to p + {}^{12}C, \quad (55)$$
$$p + {}^{12}C \to [{}^{9}B^* + {}^{4}He^*] \to p + {}^{12}C, \quad (56)$$
$$p + {}^{12}C \to [{}^{9}B^* + {}^{4}Li^*] \to p + {}^{12}C, \quad (57)$$
$$p + {}^{12}C \to [{}^{9}C^* + {}^{4}He^*] \to p + {}^{12}C, \quad (58)$$
$$p + {}^{12}C \to [{}^{9}C^* + {}^{4}Li^*] \to p + {}^{12}C, \quad (59)$$
$$p + {}^{12}C \to [{}^{9}N^* + {}^{4}He^*] \to p + {}^{12}C, \quad (60)$$
$$p + {}^{12}C \to [{}^{9}N^* + {}^{4}Li^*] \to p + {}^{12}C, \quad (61)$$

Set (19) is going to split into isobaric combinations:

$$p + {}^{12}C \to [{}^{5}Li + {}^{4}He + {}^{4}He] \to p + {}^{12}C, \quad (19)$$
$$p + {}^{12}C \to [{}^{5}Li + {}^{4}Li + {}^{4}He] \to p + {}^{12}C, \quad (62)$$
$$p + {}^{12}C \to [{}^{5}Li + {}^{4}Li + {}^{4}Li] \to p + {}^{12}C, \quad (63)$$

$$p + {}^{12}C \to [{}^{5}Be + {}^{4}He + {}^{4}He] \to p + {}^{12}C, \quad (64)$$
$$p + {}^{12}C \to [{}^{5}Be + {}^{4}Li + {}^{4}He] \to p + {}^{12}C, \quad (65)$$
$$p + {}^{12}C \to [{}^{5}Be + {}^{4}Li + {}^{4}Li] \to p + {}^{12}C, \quad (66)$$

Set (20) is going to split into isobaric combinations:

$$p + {}^{12}C \to [{}^{8}Be^* + {}^{5}Li] \to p + {}^{12}C, \quad (20)$$
$$p + {}^{12}C \to [{}^{8}Be^* + {}^{5}He] \to p + {}^{12}C, \quad (67)$$

$$p + {}^{12}C \to [{}^{8}Li^* + {}^{5}He] \to p + {}^{12}C, \quad (68)$$
$$p + {}^{12}C \to [{}^{8}Li^* + {}^{5}Li] \to p + {}^{12}C, \quad (69)$$

$$p + {}^{12}C \to [{}^{8}B^* + {}^{5}Li] \to p + {}^{12}C, \quad (70)$$
$$p + {}^{12}C \to [{}^{8}B^* + {}^{5}He] \to p + {}^{12}C, \quad (71)$$

$$p + {}^{12}C \to [{}^{8}C^* + {}^{5}Li] \to p + {}^{12}C, \quad (72)$$
$$p + {}^{12}C \to [{}^{8}C^* + {}^{5}He] \to p + {}^{12}C, \quad (73)$$

$$p + {}^{12}C \to [{}^{8}N^* + {}^{5}Li] \to p + {}^{12}C, \quad (74)$$
$$p + {}^{12}C \to [{}^{8}N^* + {}^{5}He] \to p + {}^{12}C, \quad (75)$$

**So, after calculation by *Writing I* of vertical pointers for all the combinations (13-75) no mismatches were observed – nothing or only good agreements, see Fig.8 !!!    Very promising.**

It looks like, the projectile acts by populating of a nearest state in the A=13 phase volume – an **initiated** nuclear combination, which soon will decay into primary system [$^{12}C$ g.s. + $p$] with energy of elastic $p$-peak. (The inelastic [$^{12}C^*$ + $p$] not measured.)

The populated combination comprises an "initiated" nucleus + some residual cluster, and all energy-accessible isobaric combination can play "this game" (Fig.8).

Anyway, Fig.8 in general does not look complete too.

The *Writing I* doesn't have identifiable signs of contribution of weak and/or electromagnetic forces, while the nuclear medium shows all four types of natural forces - gravitational (Mössbauer), electromagnetic, weak and strong one.

So, for the general view, *Writing I* more likely reflects work of the only strong forces.

What if one could compose another obvious expression for energy of the projectile that populate the **initiated** nuclear combination with the electromagnetic and/or weak interactions involved?

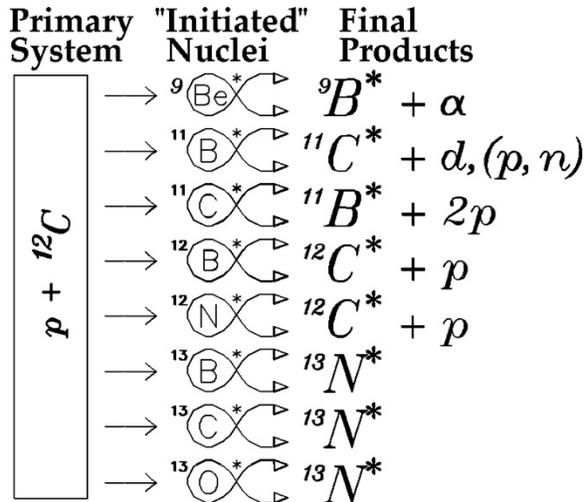
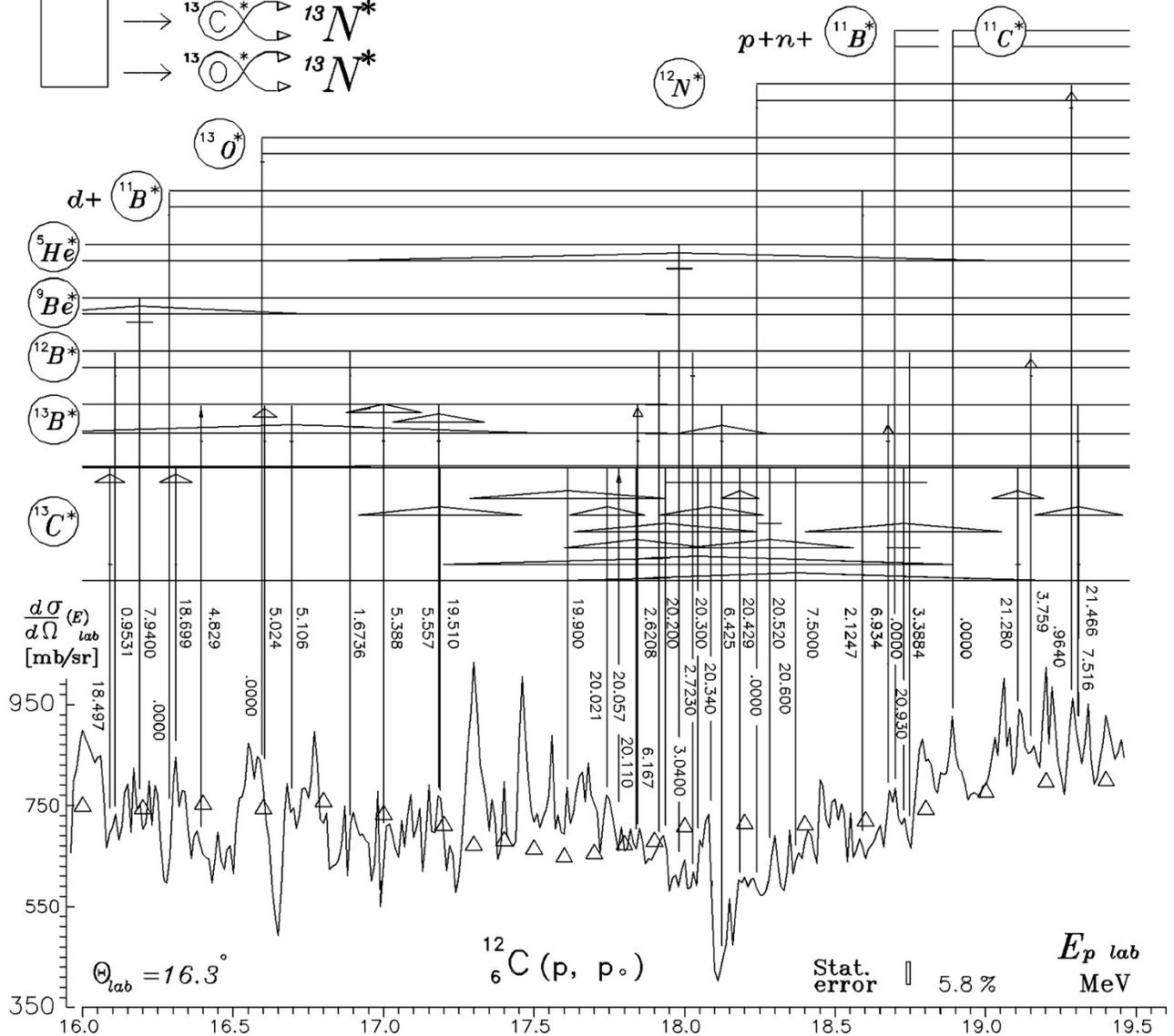

Fig. 8. Excitation function of $^{12}C(p, p_o)$ measured in the range 16 ÷ 19.5 MeV by MSS-approach at magnetic spectrograph "Apelsin" and U-150 cyclotron beam with $E_p$ =19.5 MeV with energy-spread ~200 keV.
**CIRs via strong forces -*Writing I.*** Δ - from W. W. Daehnick and R. Sherr. Phys. Rev. 133, B934

What the analog of *Writing I* for weak (WF) and/or electromagnetic (EMF) forces looks like???

Let's try *Writing II*, which has obvious terms for both the weak and electromagnetic contribution.

*Writing II*:

$$E_{Pcm} = E^*_{LIN} + E^*_{RC} + \Delta M_{FS} - \Delta M_{PS} + E_{NUC}, \quad (76)$$

$$\Delta M_{FS} = \Delta M_{FN} + \Delta m_{PP},$$
$$\Delta M_{PS} = \Delta M_{TN} + \Delta m_{BP},$$

$E_{NUC} = \Delta E_{IN} - \Delta E_{FN}$, *I* and *F*-Initiated & Final systems.

$\Delta E_{IN} = \Delta M_{IN} - Z_I \Delta m_p - N_I \Delta m_n - 0.6 Z_I(Z_I-1)\sqrt[3]{A_I}$,
$\Delta E_{FN} = \Delta M_{FN} - Z_F \Delta m_p - N_F \Delta m_n - 0.6 Z_F(Z_F-1)\sqrt[3]{A_F}$,

where $E_{Pcm}$ -is the projectile's energy in CM-frame, $E^*_{LIN}$ -is the excitation energy of the initiated nucleus (level energy), $E^*_{RC}$ -is the excitation energy of the residual cluster, $\Delta M_{FS}$-is the final system mass-excess, $\Delta M_{FN}$ -is the mass-excess of the final nucleus, $\Delta M_{IN}$ -is the mass-excess of the initiated nucleus, $\Delta m_{PP}$ -is the mass-excess of the product-particle, $\Delta M_{PS}$ -is the mass-excess of primary system, $\Delta M_{TN}$ -is the mass-excess of the target-nucleus, $\Delta m_{BP}$ -is the mass-excess of the beam-particle, *I* and *F* -Initiated and Final nuclear systems, *Z* and *N* protons and neutrons and $\Delta m_p$, $\Delta m_n - p, n$ mass-defects.

The (76) meaning is that population happens with weak and electromagnetic interactions involved too.

Application of (76) to combinations (13-75) results in Fig.12, where agreement is impressing once again.

Let's now put the Reading 0, Writing I & Writing II together on a diagram to have a general view (Fig 13).

As one can see, the general view now looks very promising, but how this could be?

The electric charge conservation is a strong law.
**What if population in the A13 phase volume happens with simultaneous production of "charge balancing" particles deeply bound in nuclear medium???**
What's known: the nuclear medium self-supports via the exchange-currents which wrap around every nucleon in the nucleus - it's a 3D-nuclear binding net of strong field that, being disturbed, makes $\pi$ / W -sounds [23-24].
Production of deeply bound $\pi$ is in study now [14-18]. But it's for heavy Pb- nuclei. How about light A13???

Well, even a neutron alone polarize vacuum so much that, by a **critical vacuum fluctuation**, neutron derives from vacuum a W which then causes neutron decay.

Mass of neutron is ~1 GeV, mass of W itself is bigger than 80 GeV.

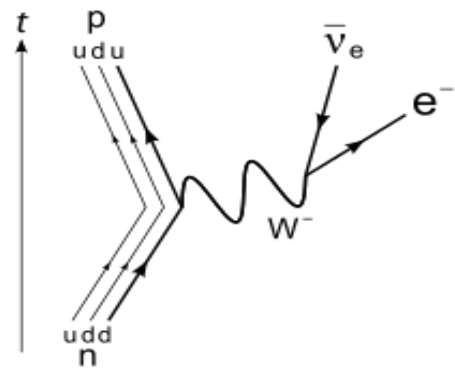

Fig.9.

Even a neutron alone can derive huge things from vacuum. Why a carbon nucleus cannot??? **It can!** "Overall" strong field of $^{12}$C is ~12 GeV. Just disturb it and it will respond with a bunch of deeply bound particles ($\pi$ or W-sounds, internally branching into leptons).

Let's make a special Feynman diagram for population of a [ $^{13}$N g.s or* + $\pi^o$] initiated system via *Reading 0*:

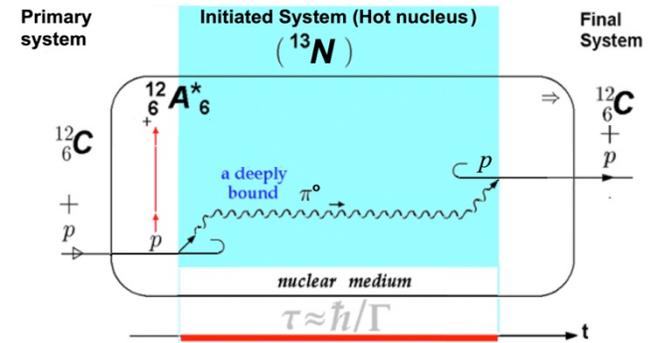

Fig. 10. Population of [$^{13}$N g.s or* +$\pi^o$] initiated system.
(Nothing via *Writing I* or *Writing II* )
Population of a $^{13}$O -initiated nucleus (in **g.s.** or excited) with a deeply bound $\pi^-$ inside is below via *Writing I*:

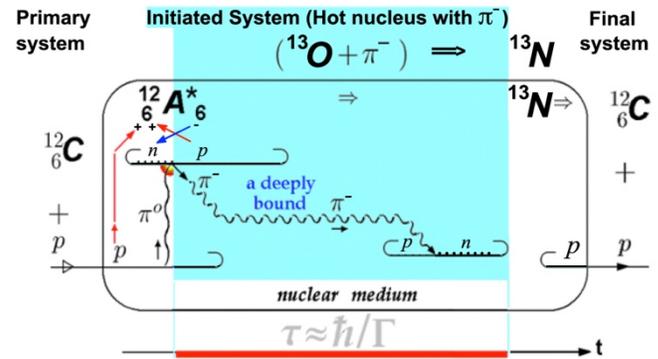

Fig. 11. Population of [$^{13}$O g.s or* +$\pi^-$] initiated system.

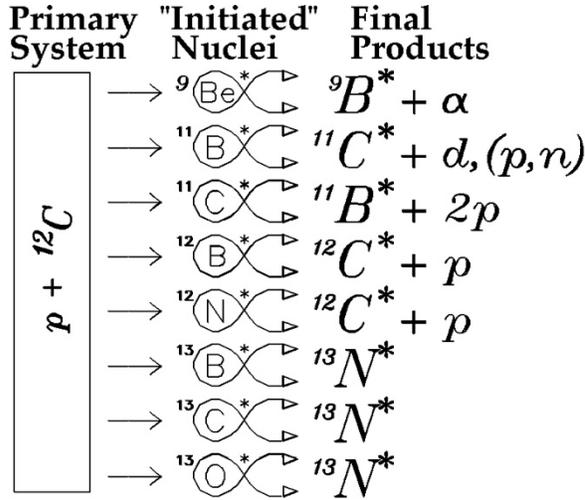
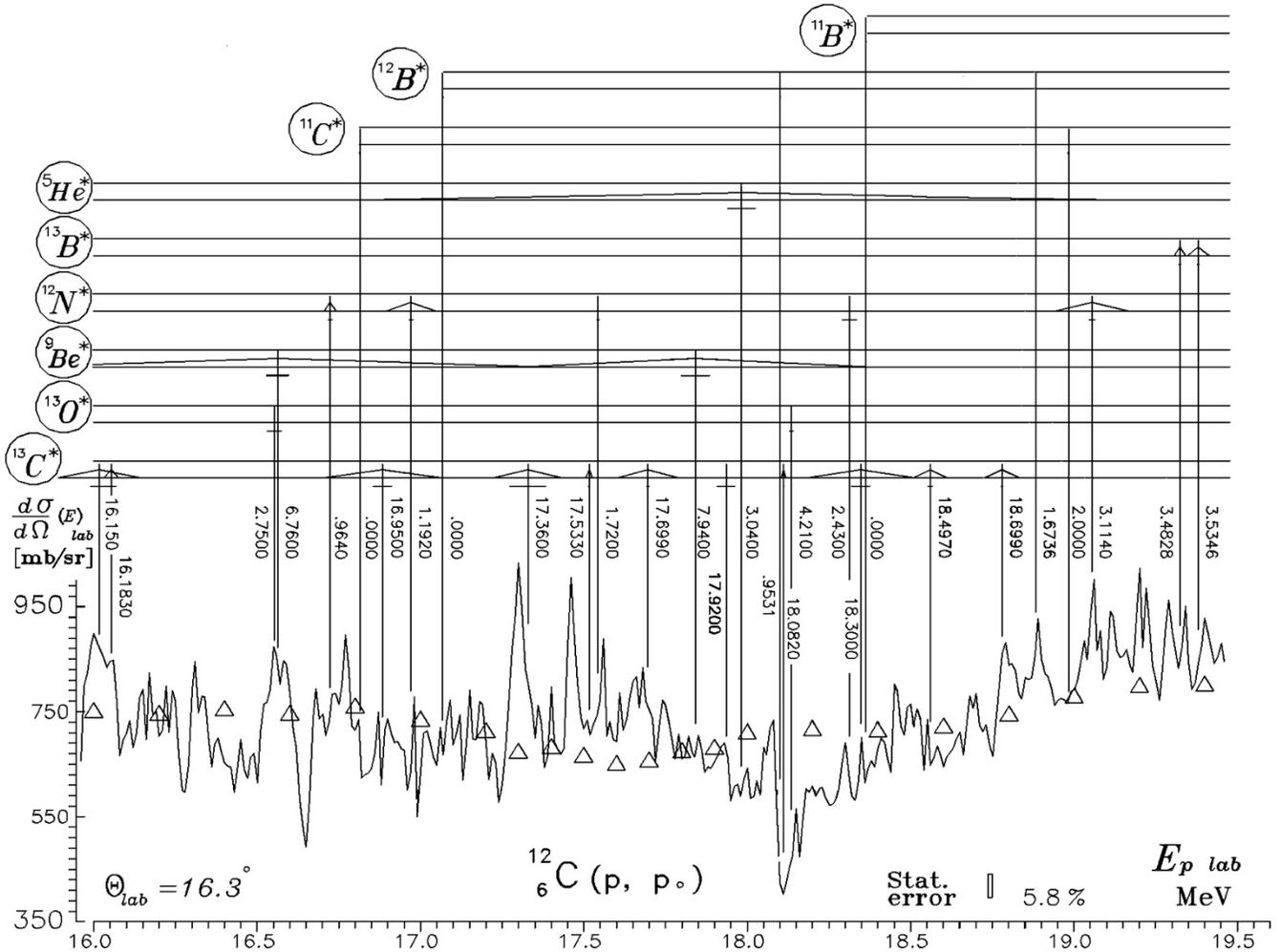

Fig. 12. Excitation function of $^{12}C(p, p_o)$ measured in the range 16 ÷ 19.5 MeV by MSS-approach at magnetic spectrograph "Apelsin" and U-150 cyclotron beam with $E_p$ =19.5 MeV with energy-spread ~200 keV.
**CIRs via weak&EM forces -*Writing II.*** Δ - from W. W. Daehnick and R. Sherr. Phys. Rev. 133, B934

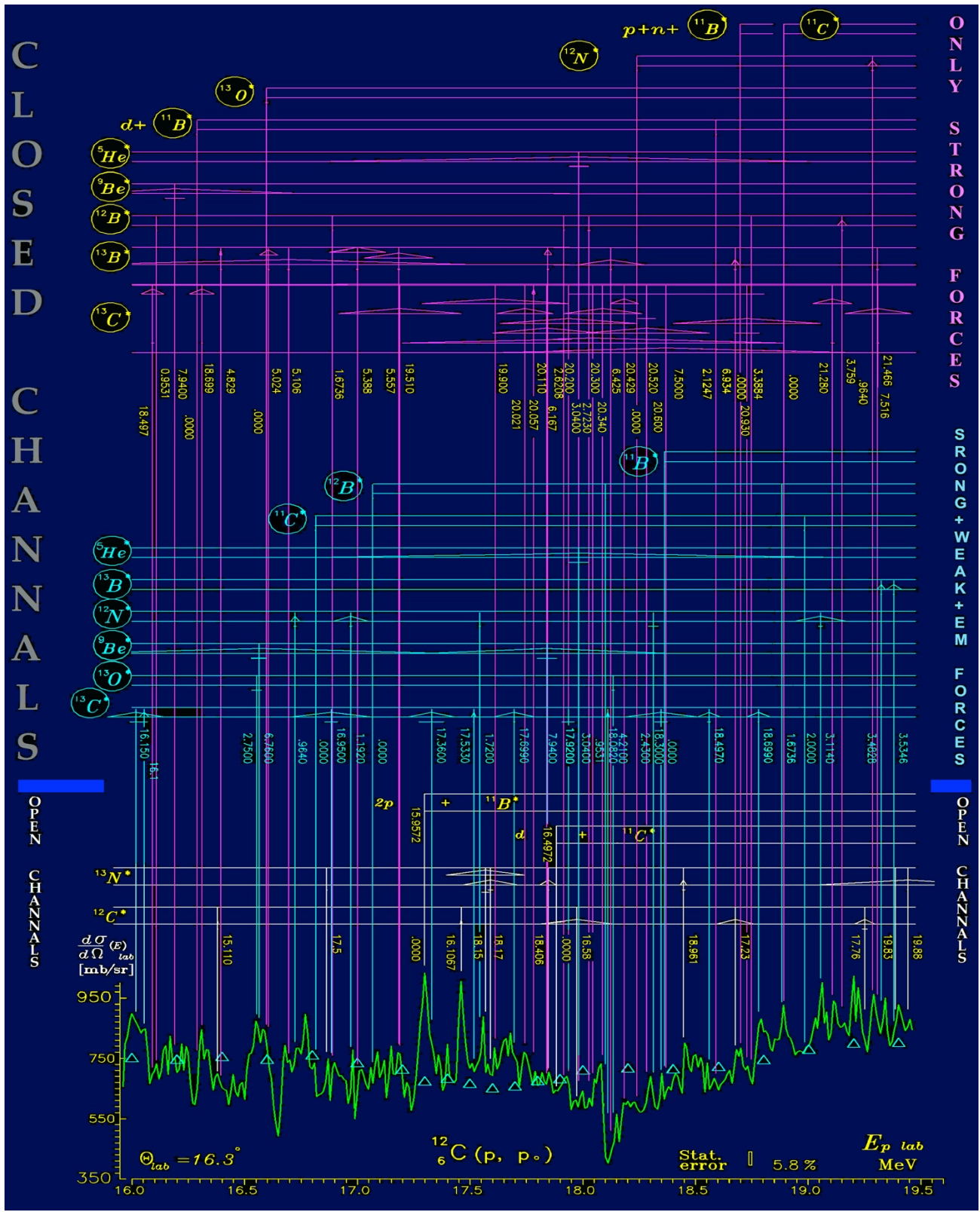

Fig. 13. The EF of $^{12}C(p, p_o)$ in the range 16 ÷ 19.5 MeV by MSS at magnetic spectrograph "Apelsin".
Possible combination isobaric resonances in A13 population according to *Reading 0*, *Writing I* and *Writing II*
Δ - from W. W. Daehnick and R. Sherr. Phys. Rev. 133, B934

Fig. 13 says one dealing with really existing thing – the resonances of combinations of isobars – the "**Combinative Isobaric Resonances**" or CIRs.

It is the reaction of a nucleus to the unexpected "energetic invasion": nuclear system accommodates the "invasion energy" using the nearest "cell" in the phase volume of arising nuclear system.

After that the quantum rules start working to get to the more stable state: the internal decay of the initiated system makes well-known output products or, in most of the cases, it decays into primary system (with highest cross-section on forward angles).

Let's make a special Feynman diagram for population by **Writing II** of an initiated system with $^{12}$**B**  ([$^{12}$**B g.s** or* + **W**$^+$] + **p**) where **W**$^+$ is normally involved in **pp** fusion, (projectile works like a **critical vacuum fluctuation** for free neutron):

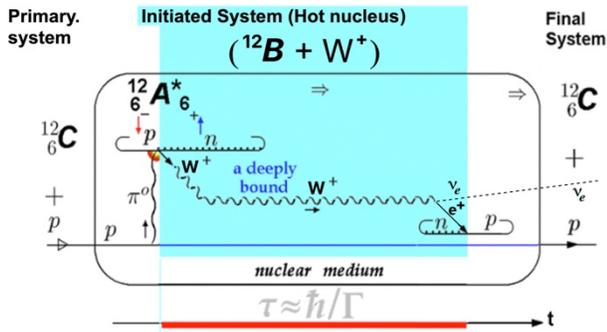

Fig.14. Population of [$^{12}$**B g.s** or*+**W**$^+$] initiated system with weak and electromagnetic forces involved.

A very important nucleus for stellar astrophysics $^{13}$**C** can be populated both ways **Writing I** or **Writing II**. Below is a Feynman diagram for population of $^{13}$**C** via only strong forces:

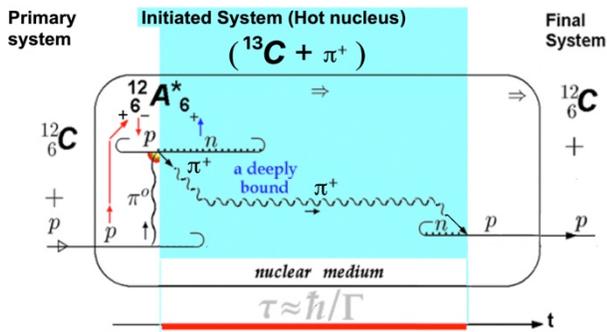

Fig.15. Population of [$^{13}$**C g.s** or*+ $\pi^+$] initiated nucleus according the **Writing I**.

Population of $^{13}$C with weak and electromagnetic forces involved could be looking like on the Fig. 16 below:

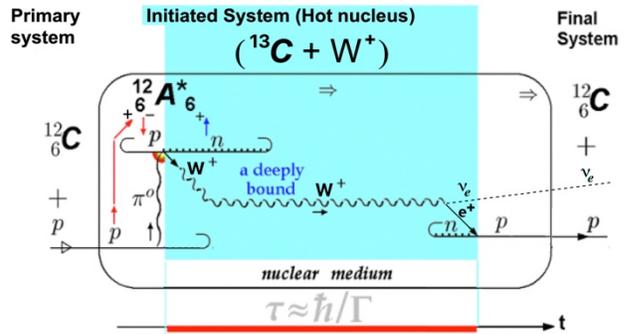

Fig.16. Population of [$^{13}$**C g.s** or*+ **W**$^+$] (**Writing II**).

It's time to say, Fig.16 could also include an intermediate chain (red brackets) of decay

$$W^+ \to [\,\mu^+ + \nu_\mu\,] \to e^+ + \nu_e \; , \qquad (77)$$

On one hand is the branching **Writing I** / **Writing II**.

But, on the other hand, there is branching in the weak and electromagnetic branch because of (77) (**Writing II**). And also, there is a number of neutrinos, what can cause widening the corresponding resonances, like it's in the β-decay (but much less: μ and π are much heavier than e and recoil doesn't change too much momentum of heavy components).

**The key question arises: how one can prove the reality of CIRs ???**
**It is easy; let's check the other precise excitation functions for presence of CIRs.**

Next preprint would be devoted to the CIRs proofs.

## 9. Acknowledgements

I would like to express my big thanks to Dr. Avas V. Khugaev and Dr. A. Avezov, to Dr. G. Kim and Dr. S. Bekbaev (all from INP, SA, RUz, Ulugbek).

A special big thanks to professor Dr. Fangil Gareev (LNR, JINR, USSR), who discussed the CIRs with me. Big thanks to Dr. Gennady A. Radiyk who many times discussed the MSS and the results with me.
Special thanks to my official opponents professor Vladimir Tokarevsky (KINR, the Ukraine) and also professor Boris Skorodumov (Kurchatov Institute, Moscow, Russia and INP SA RUz).

My sincere gratitude to BU-scientists professor James P. Miller and professor Robert M. Carey.

# References.